\documentclass[usegraphicx]{mn2e}
\begin{document}
\title[Solar system constraints on MOND]{Solar system constraints
on multi-field theories of modified
dynamics}

\author[R.H. Sanders] {R.H.~Sanders\\Kapteyn Astronomical Institute,
P.O.~Box 800,  9700 AV Groningen, The Netherlands}

 \date{received: ; accepted: }

\maketitle

\begin{abstract}
Any viable theory of modified Newtonian dynamics (MOND) as modified 
gravity is likely to require
fields in addition to the usual tensor field of General Relativity.
For these theories the MOND phenomenology emerges as an effective 
fifth force probably associated with a scalar field.  Here I consider
the constraints imposed upon such theories by solar system phenomenology,
primarily by the absence of significant deviations from inverse square 
attraction in the inner solar system as well as detectable local
preferred frame effects.  The current examples of multi-field
theories can be constructed to satisfy these constraints
and such theories lead inevitably 
to an anomalous non inverse-square force in the outer solar system.
\end{abstract}

\begin{keywords}
{Solar System, General Relativity, scalar-tensor theory}
\end {keywords}

\section{Introduction}

In modified Newtonian dynamics (MOND) it is postulated that the
effective gravitational acceleration, $g$, deviates from the Newtonian 
value, $g_N$, below a critical acceleration, $a_0$, in the sense
that $g\approx \sqrt{a_0 g_N}$ (Milgrom 1983).  That the empirically
determined value of $a_0$ ($\approx 10^{-8}$ cm/$s^2$) coincides
with $cH_0$ to within an order of magnitude was
immediately noticed by Milgrom who speculated that MOND may reflect
the influence of cosmology on local particle dynamics.  
In the context of General Relativity (GR), there is no such cosmological
influence.  This is essentially due to fact that GR embodies  
the equivalence principle in its strong form which forbids 
environmental influence on local dynamics, apart from tides.
However, if the gravitational force is partially mediated by a 
long range scalar field, as, for example, in Brans-Dicke theory,
it is no longer the case a local system is immune from cosmological
influence.  The scalar field, determined by the universal mass
distribution and its time evolution, pervades the Universe and
influences the gravitational dynamics of every subsystem.  In Brans-Dicke
theory this influence 
is evidenced by the cosmic evolution of the effective gravitational constant
(Brans \& Dicke 1961).

This suggests that MOND may have its basis in scalar-tensor theory;
indeed, the first relativistic theories proposed for MOND were 
scalar-tensor theories with non-standard aspects: the aquadratic Lagrangian
theory, AQUAL (Bekenstein \& Milgrom 1984), 
in which the scalar field Lagrangian is a general
function of the usual scalar field invariant ($F(\phi^{,\alpha}
\phi_{,\alpha})$), and phase-coupling gravity, or PCG, (Bekenstein 1988) in
which the scalar field is complex with standard Lagrangian but 
only phase coupling to matter.  Both of these early attempts contain
pathologies-- superluminal propagation or instability of the background
(Bekenstein \& Milgrom 1984, Bekenstein 1990).
Moreover, in these theories the scalar field is assumed to couple
to matter jointly with the gravitational, or
Einstein, metric in order to preserve the universality of free fall (Weak
Equivalence Principle or WEP). But if that coupling is conformal, as in
Brans-Dicke theory, 
then there is no enhanced gravitational deflection of photons due
to the scalar field.  This is in dramatic conflict with 
observations of lensing by clusters of galaxies 
(Bekenstein \& Sanders 1994).
 
The lensing contradiction led to the idea that the relation between
the Einstein and physical metrics should be more complicated than 
conformal; i.e., the so-called ``disformal transformation'' in which
certain directions are picked out for additional dilation or contraction
(Bekenstein 1993, Bekenstein \& Sanders 1994).  An initial proposal for
such a theory (Sanders 1997) invoked a non-dynamical vector field,
with only a time component in the preferred cosmological frame, 
to provide this 
additional stretching. The disformal coupling was
combined with an aquadratic Lagrangian for the
scalar field to yield the MOND phenomenology.  

However, the
non-dynamical aspect of the vector field violates general
covariance making it impossible to define a conserved energy-
momentum tensor (Lee, Lightman, \& Ni 1974).  
This problem led Bekenstein (2004) to construct a
tensor-vector-scalar theory (TeVeS) with a fully dynamical
vector field;  this theory, while yielding MOND phenomenology in
the weak field limit,
is fully covariant, produces lensing at the same
level as GR with dark matter, and possesses no obvious anomalies
(propagation of scalar waves is causal).  
In the same vein, I proposed a bi-scalar
tensor-vector theory (BSTV)
in order to provide
a cosmological origin of $a_o$ and cosmological dark matter
in the form of scalar field oscillations with wave-length 
sufficiently long to prevent clustering on the scale of galaxies
(Sanders 2005).

Thus it appears that any viable theory of MOND as modified gravity
will require fields in addition to the tensor field of GR-- a
scalar field to yield the MOND phenomenology (as a fifth force) 
and a vector field to facilitate the non-conformal coupling and adequate 
gravitational lensing. Indeed, Soussa \& Woodard (2004) have provided
an elegant no-go argument to the effect that no single metric-based theory
yielding MOND phenomenology in the weak field limit can produce
the necessary degree of gravitational lensing.  The only other
possibility for MOND as modified gravity is, then, a multi-field theory 
such as TeVeS.

The purpose of the present paper is to consider the
constraints imposed upon multi-field theories of modified dynamics
by solar system phenomenology.  The precession of the orbits
of Mercury and Icarus, as well as limits on the variation of Kepler's
constant, $GM_\odot$, between the earth and outer planets 
implies that the total force law 
within the orbit of Neptune is inverse square to high precision, 
apart from those post-Newtonian corrections
introduced by GR.  This suggests that any fifth force in
the inner solar system, in addition to preserving the WEP, 
is also precisely inverse square.  
Moreover, the absence of detectable post-Newtonian effects due to
a scalar field tied to a cosmic rest frame, i.e.,
ether-drift effects, probably constrains the magnitude of the fifth
force to be less than $10^{-4}$ that of the normal gravity force.  
But, in the context of MOND, 
the anomalous force in the Galaxy, at the neighbourhood of the
sun, would have to be comparable to the gravity force.
The transition from a weak inverse square attraction in the inner
solar system to a significant anomalous attraction at several 
thousand astronomical units (au) would seem to  
require the appearance of a non-inverse square acceleration
in the outer solar system.  

This is interesting in view of the fact that a 
deviation from inverse square attraction  
beyond 20 au is suggested on the basis
of Doppler data from the two Pioneer spacecrafts, the Pioneer
anomaly (Anderson et al.
1998, 2001). The magnitude of this apparently 
constant anomalous acceleration 
($\approx 8\times 10^{-8}$ cm/s$^2$) is tantalisingly
close to, although significantly larger than, the MOND acceleration 
(Turyshev, Nieto \& Anderson 2005).  I show here that an anomalous
acceleration is an expected, and indeed predicted, aspect of multi-field
theories of modified dynamics. This non-inverse square acceleration
appears in the outer solar system and need not be, but can be, 
as large as the observed Pioneer acceleration.
Although the discussion is general,
I illustrate this by considering current examples
of multi-field theories of MOND.

TeVeS, with Bekenstein's initial trial free function, predicts a deviation
which is too large to be consistent with both the reported constraints
on $\Delta(GM_\odot)$ and the probable limits on preferred frame
effects.  These contradictions are not fatal because the free function
of the theory can be modified to produce an anomalous force consistent
with the planetary and preferred frame constraints.  
Indeed the form of the free function required
is also consistent with that demanded by observations of
extended galaxy rotation curves which are flat beyond the visible disk.
In this case the predicted anomalous acceleration appears
beyond 100 au and is roughly $a_0/3$. 

The biscalar tensor vector theory (BSTV) is a modification of TeVeS
constructed, in part, to be
consistent with the constraints on deviations from inverse
square attraction and with the non-detection of preferred frame effects
near the earth.  It  
also predicts a constant anomalous acceleration beyond Uranus that 
depends upon the value of the scalar coupling strength. 
For values of the scalar field coupling constant below a critical
value then the constant acceleration is also $\approx a_0/3$
as in TeVeS, but for larger couplings, the
constant acceleration can be significantly larger than $a_0$
and extend within the orbit of Neptune; i.e., the theory may be tuned
to be consistent with the Pioneer effect.
If so, however, it is then inconsistent with
the reported limits on deviations from $1/r^2$ attraction  
out to the orbit of Neptune.
This is unavoidable because if the Pioneer effect is really present within
30 au, then it would be inconsistent with 
limits on variation of $GM_\odot$ between the orbits of the inner planets
and the orbit of Uranus and Neptune-- limits
derived from spacecraft ranging to these two outer planets.  Either 
this constraint on deviations from $1/r^2$ in the outer solar 
system is too stringent, which is possible (Section 4.3), 
or the reported Pioneer anomaly has a standard
explanation (not involving fundamental physics).  A more radical
possibility is that the Pioneer
effect, and hence MOND, is not due to a modification of gravity
but of the particle action (Milgrom 1994). 

\section{Multi-field theories of Modified Dynamics}

\subsection{General properties of multi-field theories}

I have emphasised that, in scalar-tensor theories of MOND, the
relation between the physical and gravitational metrics cannot be
conformal.  This condition requires the introduction of a vector field,
$A^\nu$,  that
points in the time direction in the preferred cosmological frame. 
If the
physical metric  $\tilde{g}_{\mu\nu}$ is related to the gravitational
metric $g_{\mu\nu}$ as 
$$\tilde{g}_{\mu\nu} = 
e^{-2\eta\phi}g_{\mu\nu} - 2 sinh(2\eta\phi)A_\mu A_\nu. \eqno(1)$$
then it may be shown that the scalar field enhances the deflection
of photons about a visible astronomical system
exactly as it would be by appropriately added dark matter in the context
of pure GR; i.e., relativistic and non-relativistic particles feel
the same total weak field force.
(Sanders 1997, Bekenstein 2004).  Here $\eta$ is a parameter 
describing the strength of the scalar coupling to matter
and is related to the parameter $k$ in Bekenstein's notation
($\eta^2 = k/4\pi$).  

It is useful to discuss scalar-tensor theories of modified dynamics
in the context of the Einstein frame where the scalar field 
$\phi$ may be considered
to mediate a force, $f_s$, in addition to the usual gravity force 
connected to the gravitational tensor,
the Einstein-Newton force $f_N$; that is to say, 
in this frame, particle motion
is generally non-geodesic.  In such theories the phenomenology associated 
with MOND results from this ``fifth force'' which is,
in the extragalactic domain, a non-inverse square force that dominates
in the regime of low field gradients ($f_s=\eta c^2\nabla\phi<a_0$).  

The aspect of non-inverse square attraction requires a departure from the
standard Lagrangian for scalar-tensor theories ($L_s = \phi^{,\alpha}
\phi_{,\alpha}$) either in the form
of the aquadratic theory with a non-standard scalar Lagrangian ($F(L_s)$)
or a biscalar theory where one field couples to matter and the second
determines the strength of that coupling (as in PCG).  Each of
these prescriptions may be designed to provide to a scalar force 
about a point mass of the form
$$f_s = \eta c^2\nabla\phi = {\sqrt{GMa_0}\over r} \eqno(2)$$ 
at least in the regime where
$f_s<a_0$.  The total weak field force would then be given by
$f_t = f_s + f_N$ where $f_N=GM/r^2$ is the usual Newtonian
force; clearly $f_s$ given by eq.\ 2 will dominate at accelerations
below $a_0$.

TeVeS is an aquadratic theory in disguise, with a scalar field action
that may be written as 
$$L_s = \mu\eta^2\nabla\phi\cdot\nabla\phi + V(\mu).\eqno(3)$$
This is the weak coupling limit of PCG ($\eta<<1$), the
AQUAL limit, where one may show that the kinetic term for $\mu$ vanishes.
As written here, $\mu$ is an auxiliary 
non-dynamical field and is algebraically related to $(\nabla\phi)^2$
via the potential function $V(\mu)$.  This relation, expressed in
terms of the scalar force $f_s=\eta\nabla\phi c^2$ is given by
$$\Bigl({{{f_s}}\over {a_0}}\Bigr)^2= -{l_M}^2 V'(\mu)\eqno(4)$$ where
$V'=dV/d\mu$ and $l_M=c^2/a_0$.  Here I will refer to 
-$l_m V'(\mu)$ as the free function of the theory, although this
function has, as its basis in the Lagrangian, the potential of a possibly
dynamical field, $V(\mu)$.
In the weak field static limit eq.\ 3 leads to the well known 
Bekenstein-Milgrom field equation
$$\nabla\cdot{[\mu(|f_s|/a_0)f_s}] = 4\pi G\rho. \eqno(5)$$

The function $\mu(x)$ as it appears
in eq.\ 5 does not have the same meaning as $\tilde{\mu}$
in the original MOND prescription ($f_t\tilde{\mu}(|f_t|/a_0)=f_N$)
or in the single-field Bekenstein-Milgrom non-relativistic 
theory (Bekenstein 2004).  Eq. 5
applies only to the scalar component of the force. In the context of
such multi-field theories, not all forms of $\tilde{\mu}$ are 
realisable from sensible single-valued forms of $\mu$ 
(Zhao \& Famaey 2005). 

The most obvious, and simplest, choice for the free function would
be $$\Bigl({{{f_s}}\over {a_0}}\Bigr)^2 = \mu^2\eqno(6)$$
or $\mu(x)=x$ for all $x$.  This corresponds to 
$V(\mu) = -\mu^3/(3l_M^2)$ and leads to a 
scalar force of the form of eq.\ 2 at all $r$; of course, the
Newtonian force dominates for $f_t>a_0$.
The rotation curves of spiral galaxies would be
{\it asymptotically} flat as in MOND and would satisfy a mass-rotation velocity
relation (Tully-Fisher) of the form $v^4\propto M$.  We see below, however,
that such a theory is inconsistent with the observed form of galaxy
rotation curves as well as tight constraints on deviations from
inverse square attraction in the inner solar system.

\subsection{Rotation curve constraints on fifth force theories}

A more complicated scalar field Lagrangian is 
provided by the Bekenstein free function 
$$\Bigl({{f_s}\over {a_0}}\Bigr)^2 = 
{1\over 4}\mu^2[{\eta^2\mu-2}]^2[1-\eta^2\mu]^{-1} \eqno(7)$$
(here $\mu$ as defined by eq.\ 5 differs by a factor of $\eta^2$
from Bekenstein's definition).  This yields
a scalar force illustrated by the dotted curve in Fig.\ 1 where we
see a return to $1/r^2$ attraction at high accelerations (here
$\eta=0.01$). For $f_s/a_0<10^{-4}$ this is equivalent to scalar
force dependence provided by eq.\ 6.

This free function, as well as that described by eq.\ 6, is
unacceptable in that the form of the observed rotation 
curves of spiral galaxies 
implies that the scalar force 
cannot continue to increase smoothly as $1/r$ for accelerations near $a_0$;
the resulting rotation curves decline too slowly to the asymptotically
constant value.  This has been demonstrated for the Milky
Way galaxy and for the well-studied 
spiral galaxy, NGC 3198 (Famaey \& Binney 2005), and it is generally true
(Zhao \& Famaey 2005).  Fig. 2 shows the Newtonian rotation curve
(solid curve) resulting from a spherically symmetric mass distribution of 
galaxy scale mass ($10^{11}M_\odot$), 
an exponential sphere with a length scale of 2 kpc.  The dotted
curve is the rotation curve resulting from TeVeS with 
the free function described by eq.\ 7.  The slow decline to the asymptotic
value is evident.

\begin{figure}
\resizebox{\hsize}{!}{\includegraphics{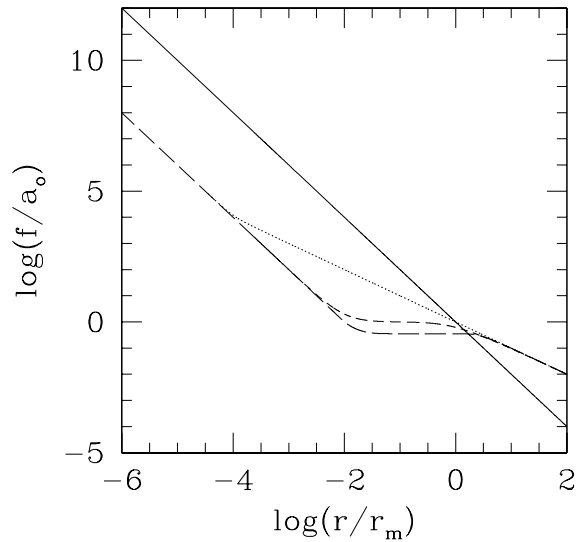}}
\caption[]{The log of the Newtonian force, $f_N$ and scalar force, $f_s$,
in units of $10^{-8}$ cm/s$^2$ plotted against the log of
the radial distance from a point mass in units of the MOND radius
($r_m=\sqrt{GM/a_0}$).  The solid curve is the Newtonian force.
The dotted curve is the scalar force for TeVeS with the
free function originally taken by Bekenstein (2005). 
The short dashed curve is the scalar force resulting from the free
function suggested by Zhao \& Famaey (2006) and the long dashed
curve is the scalar force corresponding to eq.\ 8 here. 
In all cases $\eta= 0.01$}.
\end{figure}

\begin{figure}
\resizebox{\hsize}{!}{\includegraphics{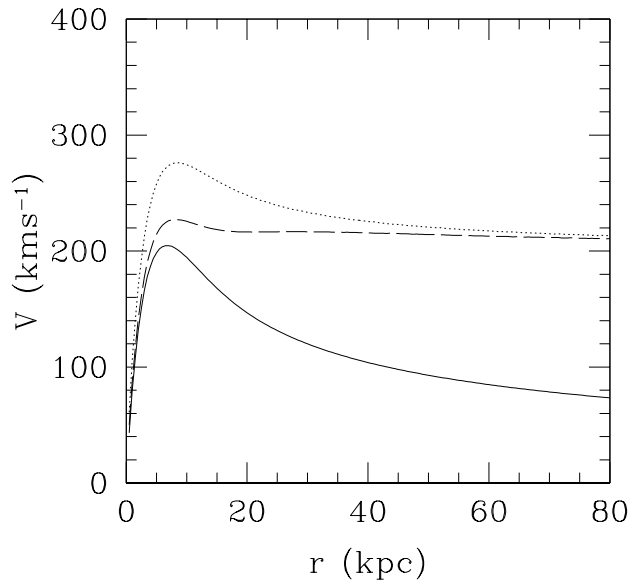}}
\caption[]{Rotation curves for a spherical galaxy resulting
from TeVeS with two alternative forms of the free function.
The density distribution is taken to be exponential with a scale
length of 2 kpc and the total mass is $10^{11}M_\odot$.
The dashed line corresponds to the original Bekenstein free function
(eq.\ 7) and the dotted curve to that of the free function considered here
(eq.\ 8).  The solid curve is the Newtonian curve.}.
\end{figure}

A free function given by 
$$\Bigl({{f_s}\over {a_0}}\Bigr)^2 = {[\mu^2 + 2\mu^4][1+4\mu^2]^{-2}}
[1-2ln(1-\eta^2\mu^2)] \eqno(8)$$ would give rise to a radial dependence
for the scalar force about a point mass of the form shown by the long
dashed curve shown in Fig.\ 1.  This is qualitatively similar to that
originally suggested for the aquadratic stratified theory (Sanders 1997)
and to the radial force dependence provided
by the free function proposed by Zhao \& Famaey (2006) (short dashed
curve).
Here we see that for total accelerations greater than $a_0$
the scalar force becomes constant, $f_s\approx a_p\approx a_0/3$ before
resuming $1/r^2$ dependence at larger total accelerations.  
The MOND interpolating function, $\tilde{\mu}$ 
(for spherical symmetry) corresponding
to the free functions of Bekenstein, Zhao-Famaey, and eq.\ 8 are 
shown in Fig. 3. 
Consistency with observed galaxy rotation curves requires a transition
to the Newtonian regime at least as rapid as that provided
by the Zhao-Famaey free function. 

The relative merit of various free functions is not the topic here;
the point is that radial dependence of scalar force must become
rather flat at accelerations larger than $a_0$ to be consistent
with galaxy rotation curves.  This is evident in Fig.\ 2 
where the rotation curve resulting from TeVeS with the free function
described by eq.\ 8 is shown by the long dashed curve; this is
more consistent with observed rotation curves which are generally flat 
beyond the visible disk.
As will be shown below, this same qualitative
behaviour is also required to meet the 
constraints on deviations from inverse square attraction in
the inner solar system while avoiding observable preferred frame 
effects.

\begin{figure}
\resizebox{\hsize}{!}{\includegraphics{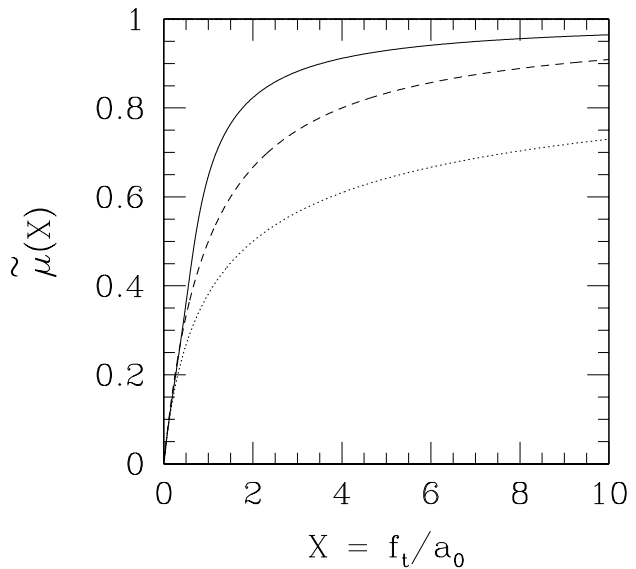}}
\caption[]{The interpolating function $\tilde{\mu}$ in the MOND 
prescription ($f_t\tilde{\mu}(|f_t|/a_0)=f_N$) corresponding to
the Bekenstein free function (eq. 7, dotted curve), the Zhao-Famaey
free function (dashed curve), and the free function considered here
(eq.\ 8, solid curve).  The indicated form is only strictly valid for
spherical symmetry, and in all cases $\eta = 0.01$
As emphasised by Zhao \& Famaey (2006) with the
Bekenstein free function the transition between the MOND and Newtonian
regimes is too gradual to yield agreement with observed rotation curves.}
\end{figure}

The BSTV theory has been designed to produce a
radial dependence of the scalar force similar in form to that given by
TeVeS with the modified free function, eq.\ 8,
(dashed curve in Fig.\ 1), and
therefore, rotation curves of the observed form (see Fig.\ 1
and 2 in Sanders 2005).  There is, however, one important difference:
Because the field determining the strength of the scalar coupling
in this theory, equivalent to $\mu$, is dynamical (unlike TeVeS 
where it is an auxiliary
field only) the value of the constant scalar acceleration near $f_s=a_0$
depends not only upon the scalar coupling strength, $\eta$, but also
upon the value of
the source mass and its distribution.  This can make an important difference
in outer solar system phenomenology.

\section{Solar system constraints}

\subsection {Planetary motion}

The most sensitive natural gravity probe in the inner
solar system is provided by the orbit of Mercury. As is well-known
General Relativity found its first
experimental success in providing a non-Newtonian explanation
of the anomalous precession of this orbit.
The precise prediction is $\Delta\theta = 43.03''$ per century, 
and the present observational result agrees with this to better than
0.05'' per century (Will 2001).  That is to say, precession resulting from
any additional non-Newtonian effect must be less than this limit.
This provides a strong constraint on long range fifth force models.

Here I will parameterise an additional non-Newtonian force in terms
of a constant acceleration $a_p$.  It is straightforward to 
demonstrate that the precession introduced by such a constant
acceleration would be given by
$$\Omega_p = {-a_p}(1-e^2)^{1\over 2}
  {\Bigl[{a\over{GM_\odot}}\Bigr]}^{1\over 2}
 \eqno(9)$$
where $e$ is the eccentricity of the orbit and $a$ is the semi-major
axis. This may be rewritten as
$$\Omega_p = 6.5(1-e^2)^{1\over 2} \Bigl({{a_p}\over{10^{-8}{\rm cm/s^2}}}
\Bigr)
 \Bigl({{10 {\rm km/s}}\over V}\Bigr)\,\,{\rm arc sec/century} \eqno(10)$$
where V is the mean orbital velocity.  If, for Mercury, 
$\Omega_p<0.05$''/century, eq.\ 10 would imply that
$a_p<4.0\times 10^{-10}$ cm/s$^2$.  That is to say, any constant anomalous
acceleration present at the distance of Mercury from the sun, must be
200 times smaller than the reported Pioneer acceleration.

The precision of measured planetary precession degrades rapidly for
the other terrestrial planets, but the asteroid, Icarus, remains
a useful probe because of its near earth passage ($a=1.08$ au) and
its high eccentricity (e=0.83).  Here, the precession predicted from
GR is 10.3 arc sec/century and the observed precession is
$\Omega_p = 9.8 \pm 0.8$ arc sec/century (Weinberg 1972).  
Taking 0.8 arc sec/century
as the limit on precession due to a constant acceleration we find, from 
eq.\ 10 that $a_p<6.3\times 10^{-8}$ cm/s$^2$,

Beyond Icarus, the tightest constraints on deviations from
$1/r^2$ attraction are provided
by limits on the variation of Kepler's constant, $K_p = GM_\odot$.
If a variation, $\Delta K_p$ is detected between two planetary orbits, 
and if this is parameterised by the presence of a constant
acceleration, $a_p$, then
$$a_p = {{\Delta K_p}\over{K_p}} {K_p}({r_2}^2-{r_1}^2)^{-1} \eqno(11)$$
where $r_1$ and $r_2$ are the distances from the sun of the closer
and more distant planets respectively (assumed to be on circular orbits).
An additional scalar force described by eq.\ 2 (in addition to the Newtonian
force) would result in a variation of Kepler's constant given by
$${{\Delta K_p}\over{K_p}} = {{r_2-r_1}\over {r_m}} \eqno(12)$$ 
($r_m = \sqrt{GM_\odot/a_0}\approx 7700$ au ).

Since the advent of interplanetary spacecrafts, the distances to the
planets are known to high accuracy.
This provides strict limits on the variation of Kepler's constant
between the earth and the planet in question,
i.e., $${{\Delta K_p}\over {K_p}} = {1\over 2}{{\Delta P}\over P}
+ {3\over 2} {{\Delta r}\over r} \eqno(13)$$ where $\Delta P/P$ is
the uncertainty in the period and $\Delta r/r$ is the uncertainty in
the solar distance.  For example, from the Viking mission to Mars
it is known that the uncertainty in the difference in the orbital
radii of Earth and Mars is less than 100 m.  Moreover, the difference
in the orbital periods between the Earth and Mars is known to 
better than 7 parts in $10^{11}$.  By eq.\ 13 this implies that
$\Delta K_p/K_p < 2\times 10^{-9}$ and hence, from eq.\ 11,
$a_p<0.1\times 10^{-8}$ for any constant acceleration present between
the orbits of the Earth and Mars (Anderson et al. 2002).
 
Similar observational limits resulting from Pioneer and Voyager
flybys constrain $\Delta K_p/K_p$ between
the inner planets and Jupiter, Uranus and Neptune to be less
than .12, 0.5, 2.0 times $10^{-6}$ respectively
(Anderson et al. 1995). The corresponding
limits on a constant acceleration are .26, .08, and .13 in units
of $10^{-8}$ cm/$s^2$.  We see that these limits for Uranus and
Neptune are inconsistent with the reported Pioneer anomaly if the
anomaly is present at distances beyond 20 au.  These results are 
summarised in Table 1.  

There is disagreement over these outer solar system constraints
(Section 4.3), but in any case, it is clear that the $1/r$
dependence of a fifth force cannot continue into the inner solar system--
certainly not to within the orbit of Mars ($r/r_m=2\times 10^{-4}$)-- 
because here the total gravitational field is so nearly inverse square.
A force law of the form of eq.\ 2 would result in 
$\Delta K_p/K_p = 7.4\times 10^{-5}$ (eq.\ 12) or 30000 times larger than the 
observed limit.  This suggests that in TeVeS
the radial dependence of scalar force must 
be quite precisely $1/r^2$ certainly within the orbit of Mars.
As we see from Fig.\ 1 this is consistent with the free function 
given by eq.\ 8 or by Zhao \& Famaey-- forms which are also consistent 
with observed galaxy rotation curves. 
Note that a theory can be constructed in which the net scalar force
vanishes within $a_0$.  This is highly contrived (involving two scalar
components, one attractive and one repulsive), but implies that, in
all that follows, one should add the condition that the scalar
force is a monotonically decreasing function of radius.

\begin{table}
\caption{Planetary constraints on a constant anomalous acceleration}
\label{symbols}
\begin{tabular}{@{}lccc}
\hline
Object & Distance (AU) & Method & $a_p$ ($10^{-8}$ cm/s$^2$)\\
\hline
Mercury & 0.39 & $\Omega_p$ & 0.04 \\
Icarus  & 1.08 & $\Omega_p$ & 6.3 \\
Mars  & 1.52 & $\Delta K_p/K_p $ & 0.1 \\
Jupiter & 5.2 & $\Delta K_p/K_p$ & 0.12 \\
Uranus & 19.2 & $\Delta K_p/K_p$ & 0.08* \\
Neptune & 30.1 & $\Delta K_p/K_p$ & 0.13* \\
\hline
\end{tabular}

\medskip
The final column is the upper limit on a constant anomalous
acceleration determined from planetary orbits via the indicated
method ($\Omega_p$, planetary precession; $\Delta K_p/K_p$,
variation of Kepler's constant).  
The constraints imposed by the orbits of 
Uranus and Neptune (marked with asterisks)
are controversial (see text), but if valid would be inconsistent
with the Pioneer anomaly as a modification of gravity.  
The distance given is the semi-major axis of the orbit..
\end{table}

\subsection{Post-Newtonian constraints}

Given that the total force must be quite precisely $1/r^2$ in the
inner solar system, it is reasonable to suppose that the scalar
force, in the high acceleration regime, is also $1/r^2$
as in Brans-Dicke theory.  Then one may ask if there is any restriction on 
the ratio of the weak field scalar to Newtonian forces in the inner solar solar
system $f_s/f_N$.
In Brans-Dicke theory, $f_s/f_N = \eta^2=1/(2\omega + 3)$  
($\omega$ is the Brans-Dicke measure of scalar coupling strength).
Because both $f_s$ and $f_N$ are inverse square in
the weak field limit, there is no restriction on the ratio
of forces, or $\omega$, from weak field phenomenology;  the 
restrictions appear at the post-Newtonian level.

In isotropic co-ordinates the metric about a point mass may be written as
$${d\tau}^2 = \Bigl[(1- {{2GM}\over{rc^2}} + 2\beta \bigl({{GM}\over r}\bigr)^2
 +...\Bigr]dt^2 - \Bigl[1+2\gamma {{GM}\over r}\Bigr]dr^2 \eqno(14)$$
where the coefficients $\gamma$ and $\beta$-- the Eddington-Robertson
parameters-- describe the lowest order
relativistic deviations from Newtonian inverse square gravity 
(post-Newtonian).  
In GR $\gamma=\beta=1$ precisely, and in Brans-Dicke theory it may be 
shown that 
$\beta =1$. In fact, this is true of any conformally coupled 
scalar-tensor theory, $\tilde{g}_{\mu\nu} = \psi(\phi)g_{\mu\nu}$
provided that $$\psi''(0) = [\psi'(0)]^2\eqno(15)$$ (see Appendix).
But, as noted in the Introduction, because of the conformal relation between
the physical and gravitational metrics, there is no enhanced deflection
of photons due to the scalar field, while non-relativistic particles 
do respond to an enhanced force.  This is reflected
in the fact that the post-Newtonian parameter $\gamma = (\omega+1)/(\omega
+2)\neq 1$ in Brans-Dicke theory.  In general, $\gamma \ne 1$ in 
in conformally coupled scalar-tensor theories.

A disformal transformation of the form of eq.\ 1 has been
discussed by Giannios (2005) in the context of TeVeS. It is equivalent 
to multiplying different components of $g_{\mu\nu}$ by
separate functions of $\phi$; i.e.,
$\tilde{g}_{tt} = \psi(\phi)g_{tt}$ and $\tilde{g}_{rr}= \chi(\phi)g_{rr}$.
In such theories, it is the case that $\gamma = 1$ if
$$\chi'(0) = -\psi'(0) \eqno(16)$$ (see Appendix).
If both conditions 15 and 16 are met and if $A^r = 0$ (i.e., the
vector does not develop a non-zero radial component), then
$\beta=\gamma=1$.  It is easy to
verify that the particular transformation 
provided by eq.\ 1, where $\psi(\phi) = exp(2\eta\phi)$ and
$\chi(\phi) = exp(-2\eta\phi)$, satisfies conditions 15 and 16.
Therefore, for any tensor-vector-scalar theory
in which the gravitational and physical metrics are related according
to eq.\ 1,
there is no restriction on $f_s/f_N$ at post-Newtonian level as
described by the standard Eddington-Robertson parameters. 
This is true of the classical stratified theories (Ni 1972), of
the stratified aquadratic theory (Sanders 1997) and of TeVeS
assuming $A^r=0$. 
These theories are consistent with a wide range of observed
phenomena from deflection of starlight by the sun to radar echo
delay.  It is the gravitational
preferred frame effects that are threatening for such theories.

\subsection{Preferred frame constraints}

Multi-field theories of MOND must contain a normalised cosmic
vector field to provide the disformal transformation. The 
direction of the vector is determined primarily by the universal
mass distribution and, in a FRW metric, points in the positive time
direction.  The equations of motion
in a gravitational field
take their simplest form in the cosmic frame where only the
time component of the vector field is non-zero.
For a frame in relative motion, such as the solar system, space
components of the vector field develop non-zero values, and this affects
the motion of particles; i.e.,
ether drift effects must appear at some level; i.e., such theories violate the
Lorentz invariance of gravitational dynamics.

Post-Newtonian preferred frame effects in conservative theories
 are quantified by two parameters,
$\alpha_1$ and $\alpha_2$ (Will \& Nordtvedt 1972).  These modify
the effective Lagrangian that describes the gravitational dynamics of N-body
systems by adding terms such as
$$\delta L_{\alpha_1} = -{{\alpha_1}\over 4}\sum_{A\ne B}{{GM_A M_B}
\over{r_{AB}\,c^2}}{\bf {V_A\cdot V_B}}\eqno(17)$$
and
$$\delta L_{\alpha_2} = {{\alpha_2}\over 4}\sum_{A\ne B}{{{GM_A M_B}\over
{r_{AB}\,c^2}}({\bf w\cdot \hat{r}_{AB}})^2} \eqno(18)$$
where ${\bf V_A}$ is the velocity of particle A with respect to
the preferred cosmic frame, ${\bf w}$ is the velocity of the inertial 
frame with respect to the cosmic frame, and $\hat{r}_{AB}$ is the
unit vector along $r_{AB}$.  A non-zero value of
 $\alpha_1$ would
lead to effects such as a polarisation of the earth-moon orbit
and is constrained to be less than $10^{-4}$ by Lunar Laser Ranging
(M\"uller, Nordtvedt \& Vokrouhlick\'y 1996).  The $\alpha_2$ term
quantifies effects such as 
periodic variation in the effective gravitational constant
(with twice the orbital or rotational frequency of the system) or
an ether drift torque acting on a spinning body.  This is constrained to
be less than $10^{-7}$ by the near-alignment of the sun's
rotational axis with that of the solar system (Nordtvedt 1987).

Calculation of the predicted values of $\alpha_1$ and $\alpha_2$
must be done for each particular theory. Here to keep the 
discussion as general
as possible, I provide estimates of the preferred frame
parameters in tensor-vector-scalar theories by heuristic arguments.

In the historical Lagrangian-based stratified theories such as that of 
Ni (1972) there
is one dynamical field, $\phi$, a non-dynamical tensor (Minkowski)
describing the background geometry, and a non-dynamical
vector field; that is to say, the gravitational force is
supposed to be mediated only by the scalar field disformally coupled
to an a priori geometry described by the Minkowski metric.  
Here it may be shown that $\alpha_2 = 0$ but
$\alpha_1 = -8$ in sharp contradiction with the LLR result, not to mention
earlier constraints on the diurnal and annual variation of the 
the gravitational constant.

In the predecessor to TeVeS, the aquadratic stratified theory 
(Sanders 1997), there are two dynamical fields, a scalar and the Einstein 
metric, in addition to a non-dynamic vector field.  That is to say,
the gravitational force is mediated not only by the scalar field, but also
by the 
Einstein metric which is locally insensitive to motion through
the cosmic frame.  Here the
preferred frame effects would appear through the contribution
of the scalar to the physical metric (ala eq.\ 1).  In the limit
where the scalar coupling, $\eta$, vanishes, the theory reduces
to GR and in GR there are no preferred frame effects.  Therefore
observational limits on preferred frame effects must place an upper
limit upon $\eta$.

This can be made more definite by noting that the equation of motion
for the scalar field, in the high acceleration limit where the
Lagrangian is standard, has the form
$$[\phi^{\alpha ;\beta}]_{;\beta} = 4\pi G\eta\tilde{T}_{\mu\nu}[g^{\mu\nu} + 
(1+e^{-4\eta\phi})  A^\mu A^\nu]\eqno(19)$$
where $\tilde{T}_{\mu\nu}$ is the usual energy momentum tensor in
the physical frame (Bekenstein 2005).
In a frame moving with velocity $w$ with respect to the cosmic frame,
the source, to order $w^2/c^2$ would take the form $4\pi G\eta
\rho[1+(w^2 + w\cdot v)/2c^2]$, but $g_{\mu\nu}$ would contain 
no such terms to post-Newtonian order.
Since $f_s = \eta\nabla \phi$, I conjecture that $\alpha_1$
is suppressed, relative to its value in pure stratified theories, by
$\approx f_s/f_N$. If so, this would constrain $f_s/f_N < 10^{-4}$

Similar arguments would apply in a theory such as TeVeS where 
all three fields, including the vector, are dynamical;
that is, we would expect post-Newtonian preferred frame effects to 
project into the solar system via the scalar field which is tied to the
cosmological frame.
There is, however, an additional effect because the vector
field contributes directly to the source of the Einstein tensor,
$G_{\mu\nu}$.  This contribution is not negligible because of the
presence, in the theory, of a Lagrangian multiplier function included to
enforce a normalisation condition on the vector field, $A_\mu A^\mu = -1$.  
The additional term in the energy-momentum tensor is then
$-\lambda A_\mu A_\nu$ which, from the vector field equation becomes
$\approx  4\pi K G_N\rho A_\mu A_\nu$ where $K$ is a new parameter 
associated with
the vector field (a coupling strength parameter), and $G_N$ is the
locally measured gravitational constant
(the primary effect of the vector field in
the weak field limit is a rescaling of the gravitational constant
with respect to its cosmological value, $G$, i.e., 
$G_N \approx G(1-K/2)^{-1}$).  
Therefore, given a Lorentz
transformation of the cosmic vector field to a moving frame we see
that the source of $G_{\mu\nu}$ contains terms proportional to 
$K G_N \rho w^2/c^2$. This would constrain K to be less than
$\alpha_1$ or $\alpha_2$ (say $K < 10^{-7}$) but would have no profound
effect on weak field phenomenology.

In summary then, we can say that the suppression of the likely
preferred frame effects such as polarisation of the
earth-moon orbit will probably require that $f_s/f_N < 10^{-4}$
This is comparable to the current reported constraints on a scalar
force in the context of Brans-Dicke theory (Bertotti et al. 2003).
But I re-emphasise, this is a heuristic argument and a proper calculation
should be done.

\section{Confrontation of multi-field theories with solar system
phenomenology}

Summarising the above discussion, we have seen, first of all,
that the total weak field gravitational force in the solar system 
is inverse square to high precision, at least within the orbit of Mars.  
This implies that any component
of the gravitational force, in addition to the Einstein-Newton force,
should also be precisely inverse square.  At the same time,
post-Newtonian preferred frame effects would seem to 
require that any additional
inverse square force due to a scalar field should be smaller than $10^{-4}$
of the Einstein-Newton force.  It is not trivial for a theory to satisfy
these two constraints.

\subsection{TeVeS}

In TeVeS, the relevant parameter, which
determines the strength of the scalar force is $\eta$; i.e.,
$f_s/f_N = \eta^2$ in the limit where $f_s>>a_0$.  The preferred frame
considerations would then require the $\eta<10^{-2}$.  
In Fig.\ 4 the dotted curve shows the
anomalous force (the non $1/r^2$ component of the total force) resulting
from Bekenstein's initial free function (eq.\ 7), with $\eta=0.01$, 
compared to the total force within the solar system (dashed curve).  
The points are the 
limits on a constant anomalous acceleration from planetary
motion discussed in section 2. The solid bar represents the
Pioneer anomalous acceleration.  Here it is obvious that the 
theory, with this choice of free function and scalar coupling strength, 
strongly violates these limits on deviations from inverse square 
attraction well into the inner solar system.

\begin{figure}
\resizebox{\hsize}{!}{\includegraphics{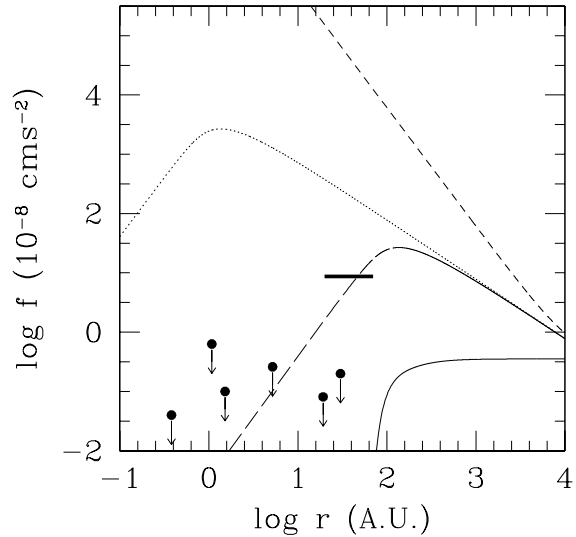}}
\caption[]{The dashed curve is the log of the 
the total force ($f_t = f_s+f_N$),
in units of $10^{-8}$ cm/s$^2$ plotted against the log of
the radial distance from the sun in astronomical units
for TeVeS.  The dotted curve is the
anomalous force (the non-inverse square force) for Bekenstein's
initial choice of free function with $\eta =0.01$ (eq.\ 7).  The long dashed
curve is the same but with $\eta=0.1$.  The solid curve is the
anomalous force resulting from the modified free function
(eq.\ 8).  The points
show the limits on a peculiar acceleration implied by planetary motion
as discussed in Section 2, and the solid bar is the Pioneer acceleration.}
\end{figure}

The deviation from inverse square attraction is less severe if the
scalar coupling constant, $\eta$ is larger.  The anomalous force resulting 
when $\eta=0.1$ is shown by the long dashed curve; this would appear
to be roughly consistent with
the constraints on deviations from inverse square within the inner 
solar system.  But then the scalar force
is only 0.01 of the Newton-Einstein force, and the theory would probably
evidence local preferred frame effects at least
100 times larger than the present limits.

The modified form of the free function (eq.\ 8) with
$\eta=0.01$, gives  
an anomalous force shown by the solid curve in Fig. 4.  It is
obvious that this is consistent both with the planetary constraints
on deviations from $1/r^2$ attraction out to Neptune
and with the avoidance of local preferred frame effects required by
$f_s/f_N<10^{-4}$, but a constant anomalous acceleration $\approx
a_0/3$ does appear beyond 100 au.  

\subsection{BSTV}

BSTV is in part 
designed to satisfy these solar system constraints but at the
expense of adding a new parameter $\epsilon>\eta$ 
(the parameter is not necessary an additional; it may be identified
with the vector coupling strength).  
Here there are two explicitly dynamical
scalar fields-- one, $\phi$, that couples to matter and the second 
$q$ which determines the strength of the coupling.  
In terms of the scalar field gradient, the quasi-static field equation is
$$\nabla\cdot[q^2 \nabla\phi] = {{8\pi G\eta\rho}\over {c^2}} \eqno(20)$$
where $q^2 \rightarrow \epsilon^2$ in the high acceleration limit; i.e.,
$q$ saturates at a small value in this limit (note that 
$\mu = q^2/2\eta^2$ with $\mu$ as defined in eqs.\ 3 and 5).  
Given that $\eta$
is the strength of the scalar field coupling (as in eq.\ 1)
and that the scalar force is $f_s = \eta c^2\nabla\phi$,
the theory is designed to yield a precisely $1/r^2$ force
in the inner solar system with 
$f_s/f_N \rightarrow 2\eta^2/\epsilon^2$ in the limit
where $f_s>>a_0$.  Thus, for this theory, the avoidance of preferred
frame effects near the earth would require that $2\eta^2/\epsilon^2
<10^{-4}$

But there is another significant difference with TeVeS.  The relation 
between the coupling strength field, $q$, and the scalar force is no
longer algebraic (as in eqs.\ 7 and 8) but is differential and given
by 
$$ {\nabla}^2 q - q\nabla\phi\cdot\nabla\phi = V_s'(q) \eqno(21)$$
where $V_s(q)$ is now an effective potential involving the
cosmic time derivative of the scalar field.  

It is instructive to view this equation in unit-less form by defining
$y=q/\eta$ and $x=r/r_m$ with $r_m = \sqrt{GM/a_0}= \sqrt{r_s l_M}$
($r_s=2GM/c^2$ is the Schwarzschild radius).  Given that $V'(q)\approx
2q^2{\dot\phi}^2/\epsilon^2$ in this regime ($q\approx \eta$), that
$a_0 = \sqrt{12}\eta|{\dot\phi}|/\epsilon$, 
and that $\nabla\phi = \eta GM/(c^2r^2q^2)$, then, in the case of spherical
symmetry, eq.\ 21 becomes
$${\eta^2{{l_M}\over{r_s}}} \nabla^2 y= {2\over{x^4y^3}}-{1\over
  {12}}y. \eqno(22)$$
From this it is obvious that, in the limit of weak coupling,
where $\eta^2 l_M/r_s <<1$, the relationship between $q^2$ and 
$\nabla\phi$ is effectively algebraic as in TeVeS (the theory approaches
its AQUAL limit).  In this case
the radial dependence of the scalar force is similar to that
shown in Fig.\  1 (dashed curves).  This 
is consistent with galaxy rotation curves and would produce an
anomalous acceleration in the solar system similar to that of
Fig.\ 4 (solid curve); i.e., it satisfies all planetary constraints
on deviations from $1/r^2$ attraction.

For a given value of $\eta$ the condition for weak coupling provides
a lower limit on the source mass.  Whenever
$$M > M_c \approx 6\eta^2 {{l_H c^2}\over G} 
\approx 5\times 10^{23}\eta^2 M_\odot \eqno(23)$$
then the weak coupling limit applies (here $l_H = c^2/H_0$ is the Hubble
radius and I have taken $l_M = 6 l_H$).  If $\eta\approx 2\times 10^{-12}$
this critical mass would correspond to a few solar masses.  In other words, 
for larger mass, the weak coupling limit applies, and the form of
the scalar force (as a function of $r/r_m$) is frozen as in TeVeS.  But for
smaller masses, the full differential equation (eq.\ 22) must be solved
and the solution depends upon the source mass (the presence of a 
critical mass in PCG was pointed out by Bekenstein 1988).

\begin{figure}
\resizebox{\hsize}{!}{\includegraphics{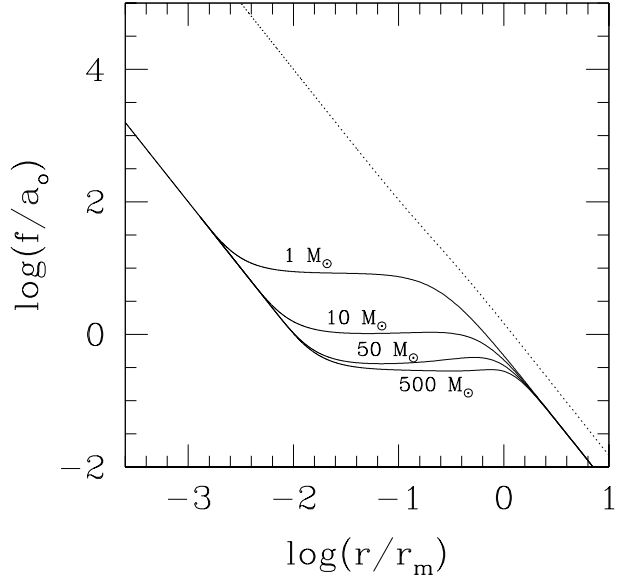}}
\caption[]{The log of the scalar force ($f_s$, solid curves) and
the total force ($f_t = f_s+f_N$, dotted curve),
in units of $10^{-8}$ cm/s$^2$ plotted against the log of
the radial distance from a point mass in units of the
MOND radius $r_m= \sqrt{GM/a_0}$ for the biscalar theory.
The various curves correspond to the indicated values of the
source mass.  The curves converge for $M>50M_\odot$ corresponding
to the weak coupling limit (the AQUAL limit) of the theory.}
\end{figure}

This is illustrated in Fig.\ 5  where we see the
scalar force as a function of the scaled radius ($r/r_m$)
for $\eta=2\times 10^{-12}$ and 
for various values of the source mass. The object is placed
in the galaxy acceleration field near the position of the sun
where $q\approx \eta$ (the scalar force is comparable to the
Newtonian force at large distance from the star).  Here, if $M>50 M_\odot$,
the solution is fixed at the weak-coupling limit.  For smaller
values of the mass, the solution, and in particular the value of
the plateau acceleration, depends upon the source mass.  If the
source mass is $1 M_\odot$, then the
plateau acceleration is $8\times 10^{-8}$ cm/s$^2$.
That is to say, unlike the weak coupling or AQUAL limit,
the constant anomalous acceleration in the outer
solar system can be significantly larger than the MOND critical
acceleration.

With $\eta = 2\times 10^{-12}$ and 
$\epsilon^2 = 2\times 10^4 \eta^2$ (Sanders 2005), the resulting anomalous 
force (solid curve) is compared to the total force (dashed curve)
in Fig.\ 6. 
Here it is evident
that, within 20 au the scalar force is 
precisely $1/r^2$ and $10^{-4}$ less than
the Newtonian force.  Moreover, the theory in this form produces
a constant anomalous force that is consistent with the Pioneer
acceleration but inconsistent
with the the reported limits the variation of Kepler's constant out to Uranus
and Neptune.  Taking the parameters of the theory to be
$\eta = 0.9\times 10^{-12}$ and $\epsilon^2 = 4\times 10^4 \eta^2$
pushes the theory back to the AQUAL limit and
produces the anomalous force shown as the dotted curve.  This is
consistent with all reported planetary constraints on inverse square 
attraction and preferred frame effects as is TeVeS with the
revised free function. In fact, it is identical in this respect to
TeVeS with the revised free function (eq.\ 8).

\begin{figure}
\resizebox{\hsize}{!}{\includegraphics{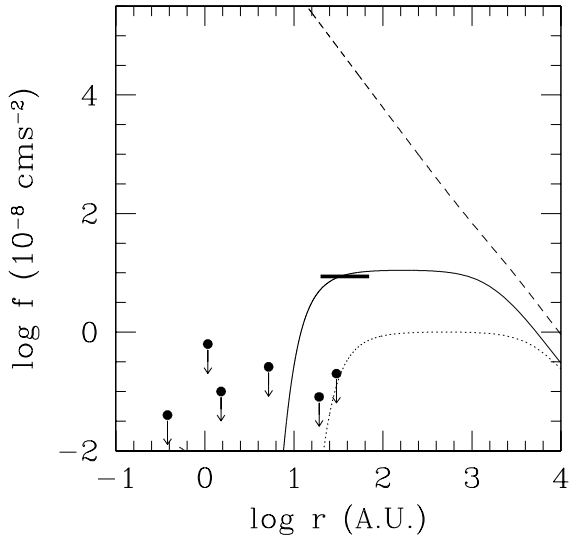}}
\caption[]{The dashed curve is the log of the total force ($f_t = f_s+f_N$),
in units of $10^{-8}$ cm/s$^2$ plotted against the log of
the radial distance from the sun in astronomical units
for the biscalar theory.  The solid curve is the anomalous force
(the non-inverse square component) for a scalar coupling
$\eta = 2\times 10^{-12}$ (with $\epsilon = 141\eta$) and 
the dotted curve is for $\eta = 0.9\times 10^{-12}$ (with 
$\epsilon=200\eta$).  As in
Fig.\ 4, the points show the planetary constraints on an anomalous 
non inverse square attraction.}
\end{figure}

\subsection{The Pioneer anomaly}

TeVeS with the free function modified to be consistent with
galaxy rotation curves (as in eq.\ 8) 
predicts a constant anomalous acceleration
beyond 100 au with magnitude $\approx 3\times 10^{-9}$ cm/s$^2$.  
The same is true of BSTV in the limit of weak scalar coupling,
$\eta<10^{-12}$.  Therefore the theories may be constructed to
be consistent both with galaxy rotation curves and with planetary
constraints on $1/r^2$ attraction within the orbit of Neptune.
In any case, an anomalous acceleration, $a_p$, in the outer solar 
system is inevitable, provided that the scalar force is a monotonically
decreasing function of radius.  Consistency with galaxy rotation
curves appears to require that $a_p\approx 0.3a_0$ as a lower limit.  
For BSTV, however,  
because the field determining the effective strength of
the scalar coupling, $q$ (or $\mu$) is dynamical, it is possible
that $a_p>a_0$ beyond 20 au while, in the outskirts of galaxies 
$a_p\approx 0.3a_0$. The same would probably be true of TeVeS
with a dynamical $\mu$.  It is
therefore tempting to identify the predicted constant acceleration
with the Pioneer anomaly.

This is problematic because, as we see in Fig.\ 5, 
the Pioneer anomaly itself is inconsistent
with reported limits on the variation of Kepler's constant out to
Uranus and Neptune (Anderson et al. 1995). However, these stated limits
may be overly stringent because they are 
based only upon single spacecraft ranging measurements to these planets,
and the formal uncertainties in the distances are
almost certainly too optimistic.
It should also be kept in mind that both Uranus and Neptune 
have not completed
a single orbit period since the advent of precise astronomical 
positioning instrumentation, and, therefore, their orbits are poorly known
(Standish 2004 and private communication 2005).

More recently, it has
been claimed that such a large anomalous acceleration, if present beyond
20 au, would lead to secular and short period signals in the orbits 
of the outer three planets-- signals large enough to
have been detected given the present levels of accuracy  (Iorio 2006). 
The opposite 
conclusion has been reached by Page, Dixon and Walen (2005) who
propose using distant asteroid orbits as a test for the Pioneer effect.
It would seem fair to conclude 
that there is an lack of agreement about
the nature of the gravitational field (as probed by planetary
orbits) in the outer solar system.  

This is an important issue.  If 
planetary motion beyond 20 au is inconsistent with the
presence of the constant Pioneer acceleration, then the Pioneer anomaly
is not due to a modification of gravity in the usual sense.  If
the planetary motion is consistent with the Pioneer anomaly, then
it remains possible that this reported constant acceleration is due
to the effect of a fifth force which becomes evident at low
accelerations, as in relativistic theories of MOND.

\section{Conclusions}

Solar system phenomenology, in particular the tight limits
on deviations from inverse square attraction and the absence to high
precision of local
preferred frame effects, places strong constraints on multi-field
theories of modified dynamics as modified gravity.   
Specifically, any fifth force
mediated by a scalar field must also be inverse square to high
precision in the inner solar system, at least within the
orbit of Mars, but smaller than about $10^{-4}$
of the Newton-Einstein force to avoid producing observable preferred
frame effects.  I re-emphasise that this constraint upon
the magnitude of the scalar force is 
only an estimate based upon
heuristic arguments; a proper calculation of the preferred
frame post-Newtonian parameters for TeVeS should be
done.  It is clear, however, that in these theories preferred frame
effects, such as a polarisation of the lunar orbit, should appear
at some level.  

In the Galaxy, at the position
of the sun, the galactic gravitational acceleration is on the order of
MOND acceleration $a_0 \approx 10^{-8}$ cm/s$^2$.  In the context of
multi-field theories of MOND this would imply that a fifth force 
acceleration is probably about
as large as the Newtonian acceleration, 
or, in terms of scalar-tensor theory,
$f_s/f_N \approx 1$. In other words, $f_s/f_N$ must grow from
0.0001 within the orbit of Neptune to about one at a distance of 800 au
where the galactic
gravitational acceleration becomes comparable to the solar attraction.  
Therefore, going outward from the sun to
the galactic environment, the scalar force must appear as an
anomalous non-inverse square acceleration (provided that the
scalar force dependence on radius is monotonic).  This effect, first noted
for the stratified aquadratic theory (Sanders 1997),
is an inevitable consequence of multi-field theories 
and is evidenced both by TeVeS and BSTV.

In order to meet the solar system constraints of precise inverse
square attraction and the absence of preferred frame effects
the scalar force must have the form demonstrated by the
dashed curves in Fig.\ 1; i.e., there must be a transition region
between $1/r$ and $1/r^2$ attraction where the acceleration due to
the scalar field is more-or-less
constant.  Therefore, not only must an anomalous acceleration appear
in the outer solar system (certainly beyond 100 au) but it must also
be, to lowest order, constant with radius between 100 and 1000 au.

Both BSTV and TeVeS
with the modified free function can provide this transition, but
there is an important difference.
In BSTV, it may be shown that
for vanishing coupling strength ($\eta\rightarrow 0$) the Laplacian
of $q$, the coupling strength field, may be neglected in the field 
equation for $q$, and the theory becomes,
in effect, an aquadratic theory as in TeVeS.  However, for finite $\eta$
the form of $q$ and hence the scalar force, in general, depends upon the
mass of the source and the coupling strength.  For $\eta<10^{-10}$
the AQUAL limit applies to galaxy scale masses but the full differential
equation must be solved for smaller masses.  The practical consequence
of this is that the plateau acceleration (where $f_s\le a_0$)
depends upon the source mass.  For a galaxy scale mass this near constant
acceleration can be $\approx a_0/3$ (as required for rotation curves)
but for a solar mass it may be near $10a_0$ if $\eta\approx 10^{-12}$.
This possibility exists for TeVeS as well if the 
auxiliary field, $\mu$, is given its own dynamics by writing a
kinetic term proportional to $\mu_{,\alpha}\mu^{,\alpha}$
into the Lagrangian (this would provide a more familiar theory).
Therefore it is possible that the the
predicted constant anomalous force could be identified with the
Pioneer anomaly.

But it is also evident from Figs.\ 4 and 6 that no theory of MOND
as modified
gravity can satisfy the reported limits on deviation from inverse 
square attraction at the orbits of Uranus and Neptune and be consistent
with the Pioneer effect if the anomalous acceleration 
really does appear at radii as small as 20 au.  This is because
the Pioneer effect itself is inconsistent with these constraints.
The constraints themselves are controversial.  Basically, the
solar gravitational field in the outer solar system is not 
well-understood, and
this calls for a reconsideration of the
orbits of the outer planets in the presence of a non-inverse square
acceleration.

It should also be noted that Milgrom (1994) has proposed
a basis for MOND as modified inertia in which the particle action
is a non-local functional of the entire particle trajectory.  This is
a completely different approach from the modified gravity theories discussed
here, and could account for the possibility that the Pioneer spacecrafts
on hyperbolic orbits feel the anomalous acceleration but the planets
on more circular orbits do not.  For this reason, it would be
of considerable interest to determine if the Pioneer anomaly first appears at
the point where a spacecraft is boosted from a bound to an unbound orbit.

The significance of the Pioneer effect should not be understated.
It may constitute the first evidence on a scale smaller than
galactic and extra-galactic that
there is more to gravity than we have supposed.
The question of
whether or not the Pioneer acceleration is a new physical effect and, 
if so, where the
anomalous acceleration first appears requires reanalysis
of the existing Pioneer data and, on the longer term, new space missions
to confirm (or not) this important result (Turyshev, Nieto \& Anderson 2004).

The possibility of such local tests is, in a sense, the ``holy grail''
of modified gravity theories (as is the direct detection of new particles
for dark matter
theories).  In this regard, Bekenstein and Magueijo (2006) have 
demonstrated that, in the context of TeVeS, MOND tidal stresses
become anomalous large near saddle points in the local solar system
where the total weak field force approaches zero, between
the earth and the sun, for example.  Space missions with
sensitive accelerometers might detect such effects.  This
would indeed be a spectacular confirmation, but 
non-detection at the predicted level would not be a falsification of 
general modified
gravity theories for MOND.  In the biscalar theory, for example,
the coupling strength field, also being dynamical, does not respond
immediately to changes in the scalar force; once in the deep Newtonian
regime, one cannot return to the MOND regime over relatively small
distances.  On the other hand,
it does appear that any theory
of MOND as modified gravity would require the presence of an
anomalous acceleration in the outer solar system (beyond 100 au) 
with a magnitude of at least a few tenths $a_0$.

I am grateful to Jacob Bekenstein, Moti Milgrom, and
Dimitrios Giannios for very useful comments
on this manuscript.  I also thank Slava Turyshev for helpful remarks
on the Pioneer anomaly and Miles Standish for comments on planetary
constraints on the solar gravitational field.

\appendix  
\section{Eddington-Robertson post-Newtonian parameters in
   theories with disformally related metrics}

Let us suppose that, in the preferred frame, the relation between 
$\tilde{g}_{tt}$ and $g_{tt}$ (the time-time components of the physical
and Einstein metrics) can be written as
$$\tilde{g}_{tt} = \psi(\phi)g_{tt}\eqno(A1)$$
where $\psi$ is a general function of $\phi$, the scalar field.
Further take the Taylor expansion 
$$\psi(\phi)=1+a\phi + {1\over 2}b\phi^2 + ...\eqno(A2)$$
where $a=\psi'(0)$ and $b=\psi''(0)$.  If the scalar field dynamics is 
described by the standard field Lagrangian, as it is for these theories
in the inner solar system (i.e., $L_s = \phi_{,\alpha}\phi^{,\alpha}$
then it is the case that $$\phi = -a{{GM}\over {r}}\eqno(A3)$$
as is usual in scalar-tensor theories.
Taking $$g_{tt} = -1 + h_{tt}\eqno(A4)$$ where, to second order
$$h_{tt} = 2{{GM}\over {r}} - 2\bigl({{GM}\over{r}}\bigr)^2 \eqno(A5)$$
Then we find, to second order 
$$\tilde{g}_{tt}=-1 + 2(1+{{a^2}\over 2}){{GM}\over{r}} - 
  2(1+a^2+{{ba^2}\over 4})\bigl({{GM}\over
   {r}}\bigr)^2. \eqno(A6)$$
Redefining the mass $M' = (1+{{a^2}\over 2})M$, we then have
$$\tilde{g}_{tt} = 1 -2{{GM'}\over{r}} - 2\bigl({{GM'}\over {r}}\bigr)^2
 \bigl[{{1+a^2+{{a^2 b}\over 4}}\over{(1+{{a^2}\over 2})^2}}\bigr].\eqno(A7)$$
By identification with eq.\ 14 we find 
$$\beta = {{1+a^2+{{a^2b}\over 4}}\over {(1+{{a^2}\over 2})^2}}.\eqno(A8)$$
The condition for $\beta=1$ is then $b=a^2$ or
 $$\psi''(0) = {\psi'(0)}^2\eqno(A9)$$
This is obviously true if $\psi(\phi) = e^{a\phi}$ which would be the
case for the particular transformation described by eq.\ 1.  
  
For a conformal transformation it is also the case that
 $\tilde{g}_{rr} = \psi(\phi) g_{rr}$.  Repeating the procedure 
above for $\tilde{g}_{rr}$ to first order, we find
$$\gamma = {{1-a^2/2}\over{1+a^2/2}}\ne 1 \eqno(A10)$$ 
(given that $a^2=2/(2\omega+3)$ this returns the usual Brans-Dicke result).

Now suppose the transformation is disformal, and of a form which generalises
eq.\ 1:
$$\tilde{g}_{\mu\nu} = u(\phi)g_{\mu\nu} - w(\phi)A_\mu A_nu \eqno(A11)$$
where $u(\phi)$ and $w(\phi)$ are unspecified functions of the scalar field.
To post-Newtonian order this is equivalent to multiplying the time-time
and space-space components of $g_{\mu\nu}$ by different functions of
the scalar field, e.g., $\tilde{g}_{tt} = \psi(\phi)g_{tt}$ and
$\tilde{g}_{rr} = \chi(\phi)g_{rr}$ where 
$$\psi(\phi) = u(\phi) + w(\phi)\eqno(A12)$$ and 
$$\chi(\phi) = u(\phi)\eqno(A13)$$
Taking $$\chi(\phi) = 1 + a'\phi \eqno(A14)$$
and repeating the above procedure for $\tilde{g}_{rr}$ we find
$$\gamma = {{1-aa'/2}\over{1+a^2/2}}. \eqno(A15)$$ 
Here $\gamma = 1$ if $a=a'$ or
$$\chi'(0) = -\psi'(0).\eqno(A16)$$ 

Rewriting conditions A9 and A16 in terms of the functions $u$ and $w$
(via eqs. 12 and 13) we find that if
$$[u'(0)+w'(0)]^2 = u''(0)+w''(0)\eqno(A17)$$
and 
$$2u'(0) = -w'(0)\eqno(A18)$$
then $\beta=\gamma=1$
It is straightforward to confirm that the particular transformation given
by eq.\ 1 satisfies these two conditions.

It is important to note that Giannios (2005) has found
two spherically symmetric static solutions for TeVeS:  one in
which $A^r=0$ everywhere (as assumed here), and a second in which 
the vector
field develops a non zero-radial component.  In this second
case $\beta\neq 0$ and depends upon the vector coupling strength
parameter and the cosmological value of the scalar field.  It is
not yet determined if both solutions are stable.

\end{document}